\documentclass[graybox]{svmult}

\usepackage{mathptmx}       
\usepackage{helvet}         
\usepackage{courier}        
\usepackage{type1cm}        
\usepackage{makeidx}         
\usepackage{graphicx}        
\usepackage{multicol}        
\usepackage[bottom]{footmisc}

\def\Tr{\mathop{\rm Tr}\nolimits}                  
\def\Res{\mathop{\rm Res}\nolimits}                
\renewcommand{\Re}{\mathop{\rm Re}\nolimits}       
\def\dir{/\kern-.7em D\,}                            
\def\lap{\Delta\,}

\def\ii{\infty}                                   
\def\Tr{\mathop{\rm Tr}\nolimits}                  
\def\Res{\mathop{\rm Res}\nolimits}                
\renewcommand{\Re}{\mathop{\rm Re}\nolimits}       
\def\dir{/\kern-.7em D\,}                            
\def\lap{\Delta\,}                                 

\def\beq{\begin{equation}}
\def\eeq{\end{equation}}

\makeindex             

\begin{document}

\title*{Generalized Zeta Function Regularization and the Multiplicative Anomaly}
\author{G. Cognola and S. Zerbini}
\institute{G. Cognola \at  Dipartimento di Fisica, Universit\`a di Trento \
and Istituto Nazionale di Fisica Nucleare - Gruppo Collegato di Trento,
Via Sommarive 14, 38100 Povo, Italia \email{cognola@science.unitn.it}
\and S. Zerbini \at Dipartimento di Fisica, Universit\`a di Trento
and Istituto Nazionale di Fisica Nucleare - Gruppo Collegato di Trento
Via Sommarive 14, 38100 Povo, Italia \email{zerbini@science.unitn.it}}
%
%
\maketitle

\abstract*{ A brief survey of the zeta function regularization and multiplicative anomaly issues when the associated zeta function of fluctuation operator is  the regular at the origin (regular case) as well as when it is  singular at the origin (singular case)  is presented. In the singular case, new results for the multiplicative anomaly are presented
.}

\section{Introduction}
\label{sec:1}

It is well known that (Euclidean) Partition function is a very important object in Relativistic Quantum Field Theory: the full propagator and all other n-point correlation functions can be computed by it. The formalism can be extended also in curved space-time \cite{buch}.
The relativistic nature of quantum  fields, namely the 
fact that an infinite number of degrees of freedom is involved, plays a crucial role. As a result,
ultraviolet divergences are present, and regularization and renormalization are necessary.    

In the one-loop approximation or in the external field approximation, one may
describe  quantum (scalar) field by means of path (Euclidean) integral and 
expressing the Euclidean partition function in terms  of functional determinants associated with differential operators. 
Namely,  the partition function is proportional to
\beq
Z_1=\left( \det L \right)^{-1/2}\,,
\end{equation}
with $L$ an elliptic self-adjoint non negative differential, the fluctuation operator. Then, the computation of 
Euclidean one-loop partition function reduces to the 
computations of functional determinants. The functional determinants 
are divergent, ultraviolet divergences are present and may be  regularized by making use of suitable  regularization.

As a simplest and illustrative example, let us consider  $\lambda \phi^4$ self-interacting scalar field. 
Let us split the quantum field as  $\phi=\Phi_0+\eta$, where  $\Phi_0$ is a classical background field. Thus  
the one-loop fluctuation operator is
\beq
L= -\lap +m^2+\frac{\lambda}{2}\Phi_0^2\,. 
\end{equation}
We recall that in  gauge theory, $A$ is singular  due to the gauge invariance and a gauge 
fixing+ ghost contributions are necessary. The one-loop quantum partition function $Z[A] $, $S_0$ being the classical action
\beq
Z[L]\simeq e^{-S_0}\int d[\eta]\,
e^{ -\frac{1}{2} \int d^4x \eta L  \eta}  
\end{equation}
reduces to a Gaussian functional integral, and as well known, it can be  
computable in terms of the real eigenvalues $\lambda_n$ of the fluctuation operator, namely $L\phi_n=\lambda_n\phi_n$. 
Since $\phi=\sum_n c_n \phi_n $, the formal 
functional measure $d[\phi]$  may be defined as  ($\mu$  arbitrary renormalization parameter)
\beq
d[\phi]=\prod_n\frac{dc_n}{\sqrt{\mu}}\,.
\end{equation}
As a consequence, the one-loop quantum "prefactor'' is
\begin{equation}
Z_1[L] = \prod_n\frac{1}{\sqrt{\mu}}
\int_{-\infty}^{\infty}dc_n
e^{-\frac{1}{2}\lambda_n c_n^2} =\left[\det(\mu^{-2}{  L})\right]^{-1/2}
\end{equation}
and the  one-loop Euclidean Effective Action reads   
\begin{equation}
\Gamma_E  =: - \log  Z = S_0+\frac{1}{2}\log ( \det \mu^{-2} {  L})\,. 
\end{equation}
What about the evalutation of the above functional determinants?
Recall the well known Schwinger argument: one starts  from the formal relation  
\begin{equation}
\log \det L=\mbox{Tr} \log L \,.
\end{equation}
Thus     
\begin{equation}
\delta \log \det L=\mbox{Tr}( L^{-1} \delta L)
\end{equation}
and consequentely one arrives at the formal expression
\begin{equation}
\left( \log \det L\right)=-\left(\int_0^\infty dt\,t^{-1} {\mbox{Tr}\,e^{-tL}}\right)\,.
\end{equation}
With regard to this expression, for large $t$, there are no problems, since  $L $  is assumed to be non negative, but for small $t$, the Heat Kernel expansion
in regular smooth and without bondary case and $D=4$, reads (see for example \cite{byts96})
\begin{equation}
\mbox{Tr}\,e^{-tL}\simeq \sum_{r=0}^\infty  A_r t^{r-2}\,.
\end{equation}
It follows that the  Schwinger representation of functional determinants is divergent a $ t=0$, and one  needs for a regularization.
One of the simplest and most useful is the dimensional regularization \cite{dowk}, which in our formulation consists in
the replacement 
\begin{equation}
t^{-1} \rightarrow \frac{t^{\varepsilon-1}}{\Gamma(1+\varepsilon)}\,.
\end{equation}
As a result, the related regularized functional determinant with $\varepsilon$ sufficiently large is
\begin{equation}
\log \det L(\varepsilon) = -\int_0^\infty dt\,\frac{t^{\varepsilon-1}}{\Gamma(1+\varepsilon)} \mbox{Tr}\,e^{-tL}=  
-\frac{\zeta(\varepsilon,L)}{\varepsilon}\,,
\end{equation}
where the generalized zeta function associated with $L$, defined for $Re s > 2$ 
\begin{equation}
\zeta(s,L)=\frac{1}{\Gamma(s)}\int_0^\infty dt\,t^{s-1} \mbox{Tr}\,e^{-tL}\,,
\end{equation}
has been introduced. In order to  be able to handle the cutoff, one makes use of the celebrated Seeley Theorem:
If $L$ is elliptic and differential operator, defined on a smooth and compact manifold, the analytic 
continuation of $\zeta(s,L)$ in the whole complex space $s$ is regular  at $s=0$. 
Making use of this  dimensional regularization, and making a Taylor expansion at $\varepsilon=0$, one arrives at
\begin{equation}
\log \det L(\varepsilon) = -\frac{1}{\varepsilon}\zeta(0,L) -\zeta'(0,L)+O(\varepsilon)\,.
\end{equation}
Thus, one obtains a justification of  the zeta-function regularized functional  determinant \cite{ray,hawk,eli0}, namely
\begin{equation}
\log \det L = -\zeta'(0,L)\,.
\end{equation}
In four dimension, the computable Seeley-de Witt coefficient $A_2=\zeta(0,L)$ controls the ultraviolet divergence,
while $\zeta'(0,A)$ gives the finite contribution, and this, in general, is difficult to evalute (see for example 
 \cite{byts96} and references therein).

\section{Multiplicative anomaly in the regular case}

In some cases, if, for example, one is dealing with a vector valued fields (charged scalar field), the $L$
becomes a matrix valued differential operator. In the evaluation of the determinant, one first computes the
algebraic one. As a consequenge, one is dealing with products of operators. A crucial point arises: the zeta-function
regularized determinants do  not satisfy the relation $ \det (AB)=\det A \det B$, or equivalently
\begin{equation} 
\ln \det
(AB)=\ln \det A+\ln \det B\,.
\end{equation}
In fact, in general,  there exists the so-called multiplicative anomaly, which may be defined as:
\begin{equation}
a(A,B)=\ln \det (AB)-\ln \det (A)-\ln \det (B)\,.  
\end{equation}
Here it is left understodd that the determinants of the two elliptic operators, $A$ and $B$
are regularized by means of the zeta-function regularization. This multiplicative anomaly has been discovered by Wodzicki (see for example \cite{eli1,eli3,guido} and references therein).

In the simple but important case in which $A$ and $B$ are two commuting invertible self-adjoint 
elliptic operators of second order, the multiplicative anomaly can be evaluated by the Wodzicki formula (a discussion can be found in  \cite{guidobook} and references therein).
\begin{equation}
a(A,B)
=\frac{1}{8} \, \mbox{res}\left[ (\ln(A B^{-1}))^2 \right]\,,
\end{equation}
where the non-commutative residue, denoted by  $ \mbox{res}$, related to a 
classical pseudo-differential operator $Q$ of order zero may defined by 
the logarithmic term in $t$ of the following generalized heat-kernel expansion
\begin{equation}
\mbox{Tr}(Qe^{-t H})= \sum_j c_j t^{(j-D)/2}-
\frac{\mbox{res}\:Q}{2} \ln t+O(t \ln t)\,,
\end{equation}
where $H$ is an elliptic non negative operator of second order, irrelevant for the evaluation of $ \mbox{res} Q$.

However, from a pratical point of view, the non-commutative residue can also be evaluated by means of the local formula
found by Wodzicki, namely
\beq
\mbox{res}\, Q =(2\pi)^{-D}\int_{M_D}dx \, \int_{|k|=1}Q_{-D}(x,k)dk\,.
\eeq
Here the homogeneity component of order $-D$ of the complete symbol appears.
Recall that a classical  pseudo-differential operator $Q$ of order zero 
has a complete symbol $ e^{ikx}Qe^{-ikx}$, admitting the following
asymptotics expansion, valid for large $|k|$ 
\beq
Q(x,k)\simeq \sum_{j=0}^\infty Q_{-j}(x,k)\,.
\eeq
In this expansion, the related  coefficients satisfy  the homogeneity property 
$Q_j(x, \lambda k)=\lambda^{-j}Q_{-j}(x,k)$.
 
\subsection{Non interacting charged boson field}

Let us consider a physical example: a free charged boson field at finite temperature $\beta=1/T$ and chemical 
potential $\mu$. The related  grand canonical partition function is standard and reads
\beq
Z_{\beta,\mu}=\int_{\phi(\tau)=\phi(\tau+\beta)}D\phi_i
e^{-\frac{1}{2}\int_0^\beta d\tau \int d^3x \phi_iA_{ij}\phi_j}\,,
\eeq
where 
\beq
 A_{ij}=\left( L_\tau+ L_3-\mu^2\right)\delta_{ij}+ 2\mu \epsilon_{ij}\sqrt{L_\tau}\,,\quad L_3=-\Delta_3+m^2\,,
\eeq
$\Delta_3$ being the Laplace operator on $R^3$, continuous
spectrum $\vec k^2$) and
$L_\tau=-\partial^2_\tau$, discrete spectrum over the Matsubara frequencies
$\omega_n^2=\frac{4\pi^2}{\beta^2}$ .
Thus,  the  grand canonical partition function may be written as 
(see, for example, \cite{eli2} and references therein)
\beq
\ln Z_{\beta,\mu}=-\ln\det \left\| A_{ik}\right\|\,.
\eeq
Now the algebraic determinant, denoted by $|A|$, can be evaluated and gives 
\beq
| A_{ik}| =( K_+K_-)\,,
\eeq
with 
\beq
K_\pm=L_3+( \sqrt L_\tau \pm i\mu)^2\,.
\eeq
However, it is easy to show that another factorization exists \cite{eli2}, i.e. 
\beq
| A_{ik}| =( L_+L_-)\,,
\eeq
where 
\beq
L_\pm=L_\tau+( \sqrt L_3 \pm \mu)^2\,.
\eeq
Now a simple calculation gives
\beq
| A_{ik}| = L_+L_-=K_+K_-\,,
\eeq
 and in both 
cases one is dealing with  the product of two pseudo-differential
operators ($\Psi$DOs), the  couple $L_+$ and $L_-$ being also
formally self-adjoint. Thus, the partition function may be written as
\beq
\ln Z_{\beta,\mu} =-\ln\det  K_+ -\ln\det K_-+a(K_+,K_-)\,,
\eeq
or as
\beq
\ln Z_{\beta,\mu}
=-\ln\det  L_+ -\ln\det L_-+a(L_+,L_-)\,.
\eeq
The evaluation of the multiplicative anomalies which appear in the above 
expressions can be  done making use of the 
Wodzicki formula and a complete agreement is found between the two expressions
 of the partition function.
Thus, if  one neglects the multiplicative anomaly, one arrives at a 
mathematical inconsistency \cite{eli2}.

\section{Multiplicative anomaly in the singular case}

However there exist cases in which the the analytic continuation of the zeta function
is not regular at $z=0$.
Recall that the usual Seeley Theorem, is based on standard heat kernel expansion (here $D=4$, and boundaryless case)   
\beq
\mbox{Tr} e^{-tL}\simeq\sum_{j=0}^\ii A_j\,t^{j-2}\,.
\eeq
As a consequence  the standard meromorphic continuation admits  only simple poles  
\beq
\zeta(s|L)=\frac{1}{\Gamma(s)}\left[ \sum_{j=0}^\infty \frac{A_j(L)}{s+j-2}+J(s) \right]\,,
\eeq
the function $J(s)$ being analytic. It follows that $\zeta(s|L)$ is regular at $s=0$ and $\zeta(0|L)=A_2(L)$,
 and $\zeta'(0|L)$ is well-defined and gives the regularized expression for $\det \ln L$. 

If we have a non standard Heat-Kernel expansion 
\beq
\mbox{Tr} e^{-tL}\simeq\sum_{j=0}^\ii A_j\,
t^{j-2}+\sum_{j=0}^\ii P_j\, \ln t\:
t^{j-2}
\eeq
namely, additional $\ln t$ terms are present, one has a generalization of Seeley result:
\beq
\zeta(s|L)=\frac{1}{\Gamma(s)}\left[ \sum_{j=0}^\infty\frac{A_j(L)}{s+j-2}
-\sum_{j=0}^\infty \frac{P_j(L)}{(s+j-2)^2}+J(s) \right] \,.
\eeq
As a consequence, double poles are present and, in general,   $\zeta(s|L)$ may have a simple pole at $s=0$.
This may happen with pseudodifferential operators  or differential  operators defined on non compact manifolds. 

Within this new contest, two issues have to be discussed. The first one is: how  $\ln \det L$ may be defined and the
second one: how   the Multiplicative Anomaly may be computed  in these singular cases?

With regard to the  first issue, the starting point is to observe that 
the functional determinat of self-adjoint operator $L$ is formally a ``divergent infinite product'' 
\beq
\prod _n \lambda_n \,,
\eeq
where $\lambda_n$ are the eingenvalues of $L$. The divergenge is present because  $L$ is a self-adjoint 
unbouned operator. 
To deals with it,  Mathematicians introduce a canonical regularization by considering the analytic continuation of 
associated zeta function $\zeta(s|L)=\sum_n \lambda^{-s}$ and then by definition 
\beq
\prod_{k=1}^{\infty}\lambda_k\equiv e^{-\left( 
\Res(\frac{\zeta(s)}{s^2})\right)_{s=0}}\,,
\eeq
where $\Res$ is the usual Cauchy residue. 

In the regular case, one Taylor expands $\zeta(s)$ at $s=0$ and one obtains
\beq
\prod_{k=1}^{\infty}\lambda_k\equiv e^{-\zeta'(0)}\,,
\eeq
in agreement with Ray-Singer-Hawking prescription. However, if one has a simple pole, namely  $\zeta(s|L)=\frac{\omega(s)}{s}$, then
\beq
\prod_{k=1}^{\infty}\lambda_k\equiv e^{-\frac{\omega''(0)}{2}}\,,
\eeq
In agreement with \cite{guido04,eli5}.
This prescription is quite general and is valid for generic singular behaviour of  $\zeta$ at $s=0$.  

What about the second issue? To our knowledge, Wodzicki approach and associated formula are valid only in the regular case.
With regard to this issue, we note that, in general, one may proceed defining the regularized functional determinant 
of the operator $L$ as regularization of a divergent product, namely
\beq
\ln \det L=-\Res\left(\frac{\zeta(s|L)}{s^2} \right)_{s=0}\,.
\eeq

\subsection{A multiplicative anomaly formula for shift operators}

Consider elliptic differential self-adjoint operators: $H=H_0+V_1$ and $H_V=H+V=H_0+V_2$ with $V=V_2-V_1$ constant 
shifts.
The main idea is to express all quantities as a function of $\zeta(s|H)$. Now, the spectral theorem gives
\beq
\zeta(s|H_V)
=\frac{1}{\Gamma(s)}\,\sum_{n=0}^\infty\,\frac{(-V)^n}{n!}\,\Gamma(s+n)\zeta(s+n|H)\,.
\eeq
\beq
\zeta(s|H\,H_V)
=\frac{1}{\Gamma(s)}\,\sum_{n=0}^\infty\,\frac{(-V)^n}{n!}\,\Gamma(s+n)\zeta(2s+n|H)\,.
\eeq
Note that here only the meromorphic continuation of $\zeta(H|s)$ appears. 

Recalling that the Multiplicative Anomaly  may be defined as
\beq
{\cal A}=\ln\det(H(H+V))-\ln \det(H+V)-\ln \det H \,,
\eeq
one has
\beq
{\cal A}={\cal A}_0 +{\cal A}_V \,,
\eeq
with
\beq
{\cal A}_0=-\Res\left( \frac{\zeta(2s|H)-2\zeta(s|H)}{s^2} \right)_{s=0}\,,
\eeq
\beq
{\cal A}_V=
-\Res \left(\frac{1}{s^2\Gamma(s)}\,\sum_{n=1}^\infty\,\frac{(-V)^n}{n!}\,\Gamma(s+n)
          \left[ \zeta(2s+n|H)-\zeta(s+n|H)\right]\right)_{s=0}\,,
\eeq

A direct computation shows that the first  term does not give any contribution to the Cauchy residue, and in the 
second term, only a {\it finite number} of terms survive, the ones corresponding to the poles for $\Re s >0$, and these are, say $n_0$. Thus, one has 
\beq
{\cal A}=-
\Res \left(\frac{1}{s^2\Gamma(s)}\,\sum_{n=1}^{n_{0}}\,\frac{(-V)^n}{n!}\,\Gamma(s+n)
          \left[ \zeta(2s+n|H)-\zeta(s+n|H)\right]\right)|_{s=0}
\,.
\eeq
This is a new result and gives a general expression for the Multiplicative  Anomaly in the case considered. It should be 
noted that the above expression   involves only the meromorphic 
continuation of $\zeta(s|H)$, and we remind that  this  follows from the Heat-Kernel  expansion of heat trace $\Tr e^{-t H}$.

Example: If one has poles of third  order, 
\beq
\zeta(s|H)=\frac{1}{\Gamma(s)}\,\sum_{j=0}^\infty \left[\frac{A_j}{s+j-2}-\frac{P_j}{(s+j-2)^2}
   +\frac{C_j}{(s+j-2)^3}+J(s)\right]\,.
\eeq
In this case $\zeta(s|H)$ has a pole of \r{second order} at $s=0$, and  we get
\beq
{\cal A}=\frac{V^2}{4}\left([A_0+(1-\gamma)P_0]
+\frac{1}{24}[10-2\pi^2-24\gamma+12\gamma^2-G]C_0\right)\,,
\eeq
 $\gamma=-\psi(1)$ being the Euler-Mascheroni constant and $G$ is a computable constant. In the standard regular case, 
all $P_j$ and $C_j$ are vanishing and
one gets the Wodzicki formula result \cite{eli1}.
\beq
{\cal A}=\frac{1}{4}\,V^2A_0=\frac{1}{96 \pi}\int d^4x (V_1-V_2)^2\,.
\eeq

Let us conclude this Section with a simple example: a vector valued massive scalar field $\vec \phi$ defined in $R \times H_3/\Gamma$, 
ultrastatic space-time with 
a non compact hyperbolic manifold with finite volume \cite{byts97}.
\beq I=\int
\left[-\frac{1}{2}\vec \phi\Delta \vec \phi+\frac{m^2 \vec \phi^2}2\right]\sqrt{g}\:d^4x
\eeq
The Heat-Kernel expansion reads
\beq
\mbox{Tr} e^{-tL}\sim\sum_{j=0}^{\ii}
\left[ A_j(L)+P_j(L)\:\ln t\right]  \:t^{j-2}
\eeq
\beq
A_0=\frac{Vol}{16 \pi^2},\quad P_0(L)=0\:,\quad \quad
P_1(L)=-\frac{Vol}{16\pi^2}\:\frac{\pi\,R}{6v_F}\:,
\eeq
\beq
P_2(L)=\frac{Vol}{16\pi^2}\:\frac{\pi\,R\,\delta^2}{6v_F}\:.
\eeq
with $v_F$ finite volume of fundamental domain of hyperbolic non compact manifold and $\delta^2=m^2+\frac{R}{6}$.
Note that $P_0=0$, thus the Multiplicative  Anomaly  is equal to the regular case. 

\section{Concluding Remarks}

The Multiplicative  Anomaly is present in dealing with functional determinants of products of differential operators. In the regular case, it is a local functional of the fields and can be computed. In the singular case, where the zeta functions are not analytic at $s=0$, we have shown that it is still a local functional and we have provided a formula fot its evaluation. 

Within one-loop physics, apparently no new physics seems to be associated with Multiplicative Anomaly, also in the presence  of generalized zeta-function regularization. However its inclusion is necessary for mathematical consistency: charged scalar field at finite temperature is an example.

Furthermore, dealing with spinor fields, one has the Euclidean massive Dirac operator. 
\beq
K=p_\mu \gamma^\mu +iM=A+iM\,,\quad  A^+=A=p_\mu \gamma^\mu \,. 
\eeq
Problem: How to evaluate  $\ln \det A$ ? In $D=4$, with $L=A^2$ being the  spinorial Laplace operator in curved space, one has
\cite{eli4}
\beq
\ln \det K=\frac{1}{2}\ln \det(L+M^2)+i\frac{\pi}{2}\zeta(0|L+M^2)+c_1M^2 A_1(L)+c_2M^4 A_0(L)\,.
\eeq
where $A_0$ and $A_1$ are the associated Seeley-de Witt coefficients. 
Two remarks: first the last term contains $A_1(L)\equiv R$, Ricci scalar, this is Sakarov induced gravity idea 
\cite{eli3}.  
Second, this last term may be interpreted as multiplicative anomaly contribution \cite{eli4}. With regard to this last issue, recently , I. Shapiro and other have reported a non trivial non-local M. A . in Quantum ED in curved space-time 
\cite{ilia10}.

\begin{acknowledgement}
We would like to thank our friend Emilio Elizalde for fruitful colloboration on this interesting issue in all these years. 
\end{acknowledgement}

%
%
%

\end{document}